\documentclass[review]{elsarticle}

\usepackage{lineno,hyperref}
\usepackage{subfigure}
\usepackage{amsmath, amssymb}
\usepackage{amsbsy}
\usepackage{mathptmx}
\usepackage{bm}
\usepackage{makecell}
\usepackage[linesnumbered,lined,boxed,commentsnumbered]{algorithm2e}
\usepackage{color}
\usepackage{latexsym}
\DeclareMathOperator{\sign}{sign}
\modulolinenumbers[5]
\usepackage[margin=1. in]{geometry}
\journal{X}

\begin{document}

\begin{frontmatter}

\title{On the numerical overshoots of shock-capturing schemes}

\author[1]{Huaibao Zhang}
\author[3]{Fan Zhang\corref{cor1}}
\ead{fan.zhang@kuleuven.be, fan.zhang@mail.be}
\cortext[cor1]{Corresponding author.}
\address[1]{School of Physics, Sun Yat-sen University, Guangzhou 510006, China}
\address[3]{Centre for mathematical Plasma-Astrophysics, Department of Mathematics, KU Leuven, Celestijnenlaan 200B, 3001 Leuven, Belgium}

 \begin{abstract}
This note introduces a simple metric for benchmarking shock-capturing schemes. This metric is especially focused on the shock-capturing overshoots, which may undermine the robustness of numerical simulations, as well as the reliability of numerical results. The idea is to numerically solve the model linear advection equation with an initial condition of a square wave characterized with different wavenumbers. After one step of temporal evolution, the exact numerical overshoot error can be easily determined and shown as a function of the CFL number and the reduced wavenumber. With the overshoot error quantified by the present metric, a number of representative shock-capturing schemes are analyzed accordingly, and several findings including the amplitude of overshoots non-monotonously varying with the CFL number, and the amplitude of overshoots significantly depending on the reduced wavenumber of the square waves (discontinuities), are newly discovered, which are not before.
\end{abstract}


\end{frontmatter}

\section{Introduction} \label{sec:intro}
Numerical simulations of compressible flows are a highly challenging topic since shock waves, which cause strong density, velocity, and pressure discontinuities, need to be tackled by using robust \cite{Harten1984}, accurate \cite{Harten1987,Liu1994} and efficient \cite{Zhang2020a,Xu2021} shock-capturing schemes, the development of which, however, is still not trivial. This topic becomes even more challenging when it comes to high-order shock-capturing schemes, since for which, the satisfaction of high numerical resolution and strong computational robustness are contradictory. More specifically, achieving high-resolution usually relies on high-order polynomials for the spatial approximation, but the high-order polynomials tend to yield oscillatory solutions as they are crossed by discontinuities. The oscillation, also known as the Gibbs phenomenon, cannot be reduced in amplitude through grid refinement.

The past decades have seen the successful development of several high-resolution shock-capturing schemes~\cite{Harten1984,Harten1987,Liu1994,Jiang1996,Borges2008}. For instance, the total variation diminishing (TVD) scheme \cite{Harten1984} is a canonical shock-capturing scheme. It enforces the condition of non-increase of the total variation in time, usually by using flux or slope limiters, capable of achieving oscillation-free shock-capturing solutions. However, it is well known that the TVD scheme suffers from order reduction at the smooth extrema of the solution. Uniform high-order accuracy can be achieved by the essentially non-oscillatory (ENO) scheme which uses adaptive stencils to avoid, or at least minimize the reconstruction crossed by discontinuities, thus reducing the oscillation or overshoot essentially.
An important improvement of the ENO scheme, the weighted essentially non-oscillatory (WENO) scheme \cite{Liu1994}, uses a convex-combination procedure of reconstructions carried out over a given amount of candidate stencils, and the shock-capturing performance is highly dependent on the associated nonlinear weights which in essence represent the relative local smoothness of the solution over each candidate stencil.
WENO schemes \cite{Liu1994,Jiang1996,Borges2008} maintain the ENO property and significantly improve the overall resolution. However, ENO/WENO schemes cannot completely remove the numerical oscillation or overshoot in shock-capturing computations \cite{Zhao2019}.
The oscillation not just contaminates the numerical solution, but also causes the computations to blow up at critical conditions.
To increase the robustness of WENO schemes, great efforts have been made, and one improved version, termed monotonicity-preserving WENO (MPWENO) scheme, is then proposed, which is capable of running severe test cases involving blast waves and/or near-vacuum \cite{Balsara2000,Hu2013}.

It is not the intention in this note to improve the performance of the aforementioned shock-capturing schemes. Rather, we restrict ourselves to a suitable metric of assessing the shock-capturing error, since as we have investigated in the literature that the overshoot error near the shock yielded by the numerical scheme is still mostly evaluated based on a common case-by-case fashion, through visualized investigation and comparison among numerical results, rather than a quantitative analysis. Very recently, a quantitative metric for the shock-capturing error has been given by Zhao et al. \cite{Zhao2019}, showing that the ENO and WENO schemes may yield significant over-amplification of incoming waves at discontinuities. This phenomenon occurs possibly due to the activation of linearly unstable stencils in the vicinity of the discontinuity, and does surprisingly within a certain range of wavenumber space. The traditional evaluation approach relying on simple shock simulations, however, may fail to trigger the over-amplification behavior as the wavenumber falls out of the target region, and thus cannot provide very useful information.

Despite the successful work of Zhao et al. \cite{Zhao2019}, we intend to move further in the investigation of shock-capturing error metric by evaluating the overshoot errors of the aforementioned shock-capturing schemes in quantitative analysis. For that purpose, a model linear advection equation is carefully selected instead and numerically solved by the use of these shock-capturing schemes. The initial condition is specified by a square wave of different wavenumbers, in which the flow variable jump mimics the shock discontinuity. The numerical overshoot error after one step of time integration is then exactly computed by comparing the numerical solution against the analytical one. Through the proposed metric, at least three problems are interesting to investigate.
First of all, the influence of the CFL number on the overshoot error, which plays a key role in the robustness of numerical simulation, waits to be unveiled as the overall perspective of the quantified overshoot error versus the CFL number is plotted. Secondly, the numerical behavior of capturing multiple closely located shock waves is considered. The third problem involves the overshoot error characteristics.

The remainder of this work is organized as follows. The tool of quantitative analysis is introduced in section \ref{eq:method}, and then in section \ref{sec:results} several canonical shock-capturing schemes are investigated by the present metric. Eventually, the concluding remarks are given in the last section.

\section{The numerical techniques} \label{eq:method}

\subsection{The initial-boundary value  problem (IVP)} \label{sec:ivp}

We consider the IVP of one-dimensional wave propagation in an
unbounded domain, governed by the scalar linear advection equation, i.e.
\begin{equation} \label{eq:hcl}
\frac{{\partial u}}{{\partial t}} + a\frac{{\partial u}}{{\partial x}} =0,
\end{equation}
\noindent where the constant $a$ denotes the characteristic velocity. In  practice, the spatial discretization of Eq.\eqref{eq:hcl} is performed on an equally spaced grid, resulting in a system of  ordinary differential equations (ODE)
\begin{equation} \label{eq:dis}
\frac{{d u_i}}{{d t}}=-a \frac{{\partial u}}{{\partial x}}|_{x=x_i}, \quad 0 \le x_i \le L, \quad i=1, \cdots, N,
\end{equation}
\noindent and the periodic boundary condition is given at the left ($x=0$) and right ($x=L$) boundaries to model the unbounded domain.

The fundamental idea of the general framework for the evaluation of shock-capturing error in this work is defining the initial field by a series of square waves to mimic the presence of discontinuities, which is described by
\begin{equation}
 u(x,T_0)=0.5\sign\left(\sin(2\pi x/\lambda_n)\right),
\end{equation}
\noindent where $T_0=0$, and $\lambda_n$ is the wave length, defined as
\begin{equation}
\lambda_n = L/n, \quad n = 1,\cdots ,N/2.
\end{equation}
\noindent
Governed by the Eq.~\eqref{eq:hcl}, the exact solution after a time $T$ is just a translation of the initial condition with a distance of $aT$, given in the form of
\begin{equation}
u(x,T)= u(x-a T, T_0).
\end{equation}

The reduced wavenumber is introduced to characterize the shock-capturing overshoot error for the rest of the analysis, which is defined as
\begin{equation}
\varphi=\omega \Delta x=(2\pi/\lambda)\Delta x,
\end{equation}
\noindent where $\omega=2\pi/\lambda$ is the wavenumber.

We set $a=1$, without loss of generality, $N=500$ and $L=1$ for the following numerical simulations.

\subsection{Shock-capturing schemes}
The derivative term on the right-hand side of Eq.~\eqref{eq:dis} can be approximated by a conservative finite difference formula
\begin{equation}\label{eq:semi}
\frac{{d u_i}}{{d t}}=- \frac{1}{\Delta x}(h_{i+1/2}-h_{i-1/2}),
\end{equation}
\noindent where the flux function $h_{i\pm 1/2}$ at half points can be implicitly defined by
 \begin{equation}\label{eq:flux}
f(x)=\frac{1}{\Delta x}\int_{x-\Delta x /2}^{x+\Delta x /2} h(\xi)d\xi,
\end{equation}

 \noindent and  then a
 semi-discretized form can be   written as
\begin{equation}\label{eq:app}
\frac{{d u_i}}{{d t}}\approx - \frac{1}{\Delta x}(\hat{f}_{i+1/2}-\hat{f}_{i-1/2}),
\end{equation}
\noindent  where
\begin{equation} \label{eq:nonlinear}
\hat{f}_{i+ 1/2}=\hat{f}\left(f_{i-l},\cdots,f_{i+r}\right),
\end{equation}
\noindent and $\hat{f}_{i-1/2}$ can be easily obtained symmetrically.
Shock-capturing schemes are used for the approximation of $\hat{f}$ when the flowfield contains discontinuous solutions such as shock waves or contact discontinuities. $l$ and $r$ define the widths of the stencil required for the evaluation of the flux term $\hat{f}$~ by the specific shock-capturing scheme.

Among the shock-capturing schemes in the literature, the classical TVD schemes \cite{Harten1984,Sweby1984} and the ENO/WENO schemes \cite{Harten1984,Harten1987,Liu1994,Jiang1996,Borges2008} are considered for the investigation in this work. TVD schemes are known as oscillation-free in one-dimensional simulations. In particular, we choose the second-order TVD schemes~\cite{Sweby1984} with the minmod (MM) limiter and superbee (SB) limiter among the TVD family, since the MM limiter is considered to be the most dissipative limiter, and the SB limiter has a more compressive nature than the other limiters.




\subsection{The error metric}

Rather than the statistics of the overall numerical error, we are more concerned on the overshoot error near the discontinuities since it can lead to negative pressure or density in gas dynamic simulation, and the computation blows up as a result. Only the exaggerated amplitude or overshoot is therefore counted. It is also our attempt to exclude the inherent dissipation effect of the specific numerical scheme to the numerical solution.
According to the work of Deng et al. \cite{Deng2019}, numerical dissipation in long-time simulations can inevitably smooth out all the discontinuities and smooth waves, even for the high-order WENO schemes.
Here, only one step of temporal evolution is marched with a time $T$ which is related to the $\text{CFL} \in (0,0.9]$.
The time integration is performed by using the third-order strongly stable Runge-Kutta method \cite{Gottlieb2001}. The numerical solution obtained can be formulated as
\begin{equation}
\begin{split}
\hat{u}(x,T)&=u(x,T)+u_{\epsilon}(x,T) ,\\
\end{split}
\end{equation}
\noindent where $u_{\epsilon}(x,T)$ is the error function.
We further define the numerical overshoot by
 \begin{equation}
u_{\epsilon}^{\text{overshoot}}(x)=
\begin{cases}
\begin{matrix}
|u_{\epsilon}(x, T)|, & \text{if} \quad |\hat{u}(x,T)|>0.5, \\
0, & \text{otherwise}.
\end{matrix}
\end{cases}
\end{equation}
\noindent

By this definition, we exclude the numerical error caused by the dissipative effect from our discussion. For each combination of $T$ and $\lambda_n$, only the maximum overshoot, i.e., $\max(u_{\epsilon}^{\text{overshoot}}(x))$, is shown in the next section.



\section{Results and discussions} \label{sec:results}%
As mentioned above, we only consider several representative shock-capturing schemes, including (I) the second-order TVD schemes \cite{Osher1984,Sweby1984} with the minmod (MM) and superbee (SB) flux limiters; (II) the ENO schemes \cite{Shu1989} having third-order and fifth-order accuracy; (III) the WENO-JS schemes having fifth-order \cite{Jiang1996} and seventh-order \cite{Balsara2000} accuracy;
  (IV) the WENO-Z schemes having fifth-order \cite{Borges2008} and seventh-order \cite{Don2013} accuracy; and
  (V) the  MPWENO schemes having fifth-order and seventh-order accuracy  \cite{Balsara2000}.

\begin{figure}[htbp]
 \centering
 \subfigure[\label{fig:minmod}{minmod}]{
 \includegraphics[width=0.48\textwidth]{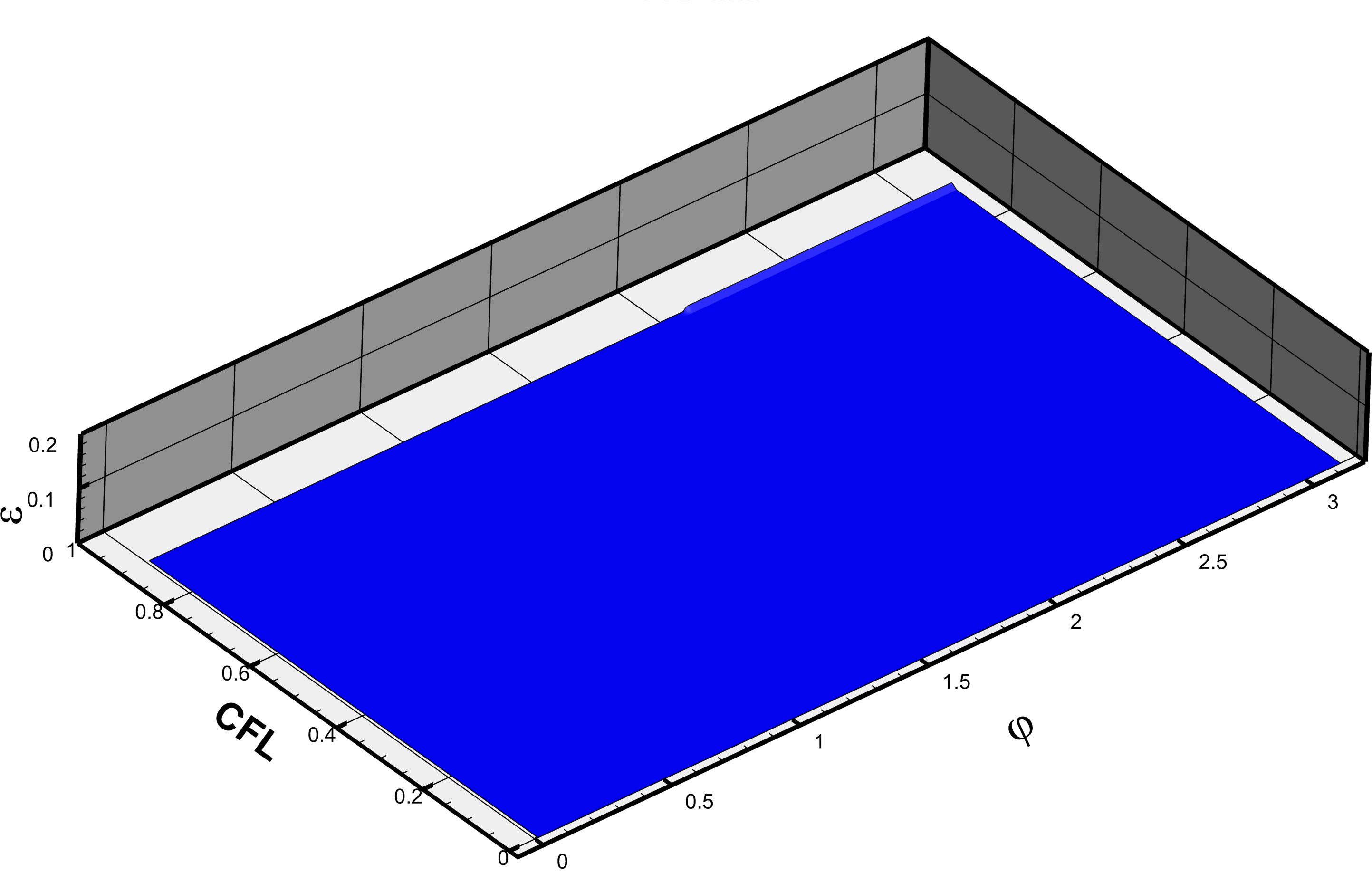}
 }
 \subfigure[\label{fig:superbee}{superbee}]{
 \includegraphics[width=0.48\textwidth]{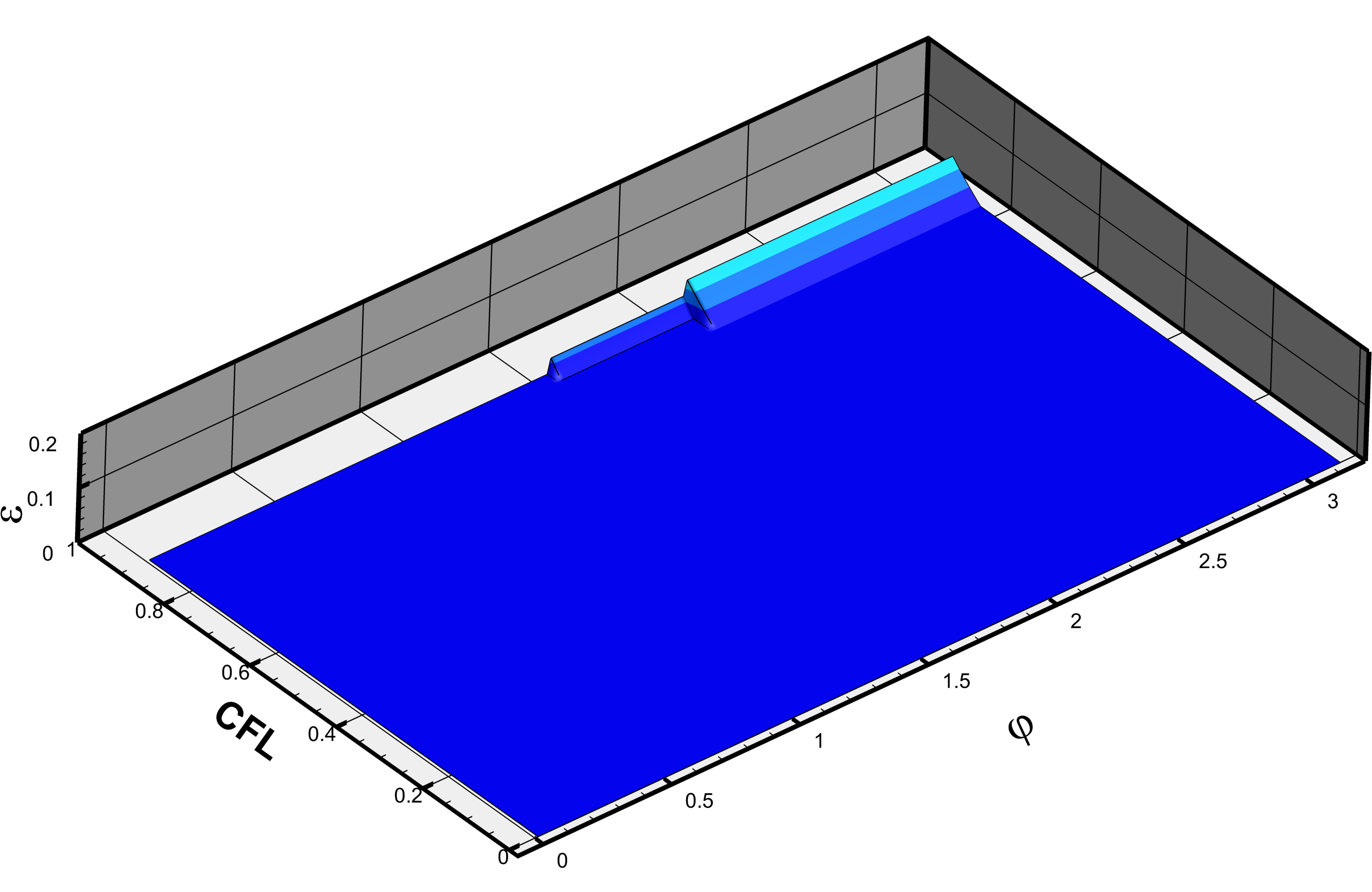}
 }
 \caption{Numerical overshoots of TVD schemes.}
 \label{fig:TVD}
\end{figure}

As shown in Fig.~\ref{fig:minmod}, the TVD scheme with the MM limiter guarantees overshoot-free results even using a large CFL number, while using the SB limiter, it produces minor overshoots as the wavenumber and CFL number are both large, as shown in Fig.~\ref{fig:superbee}.
In general, these two limiters can be considered to be able to eliminate the oscillation near discontinuities, and thus, the other TVD limiters, which are more restrictive in the limiting constraint than the SB limiter, are expected to suppress the oscillation as well.

\begin{figure}[htbp]
 \centering
 \subfigure[\label{fig:ENO3}{ENO3}]{
 \includegraphics[width=0.48\textwidth]{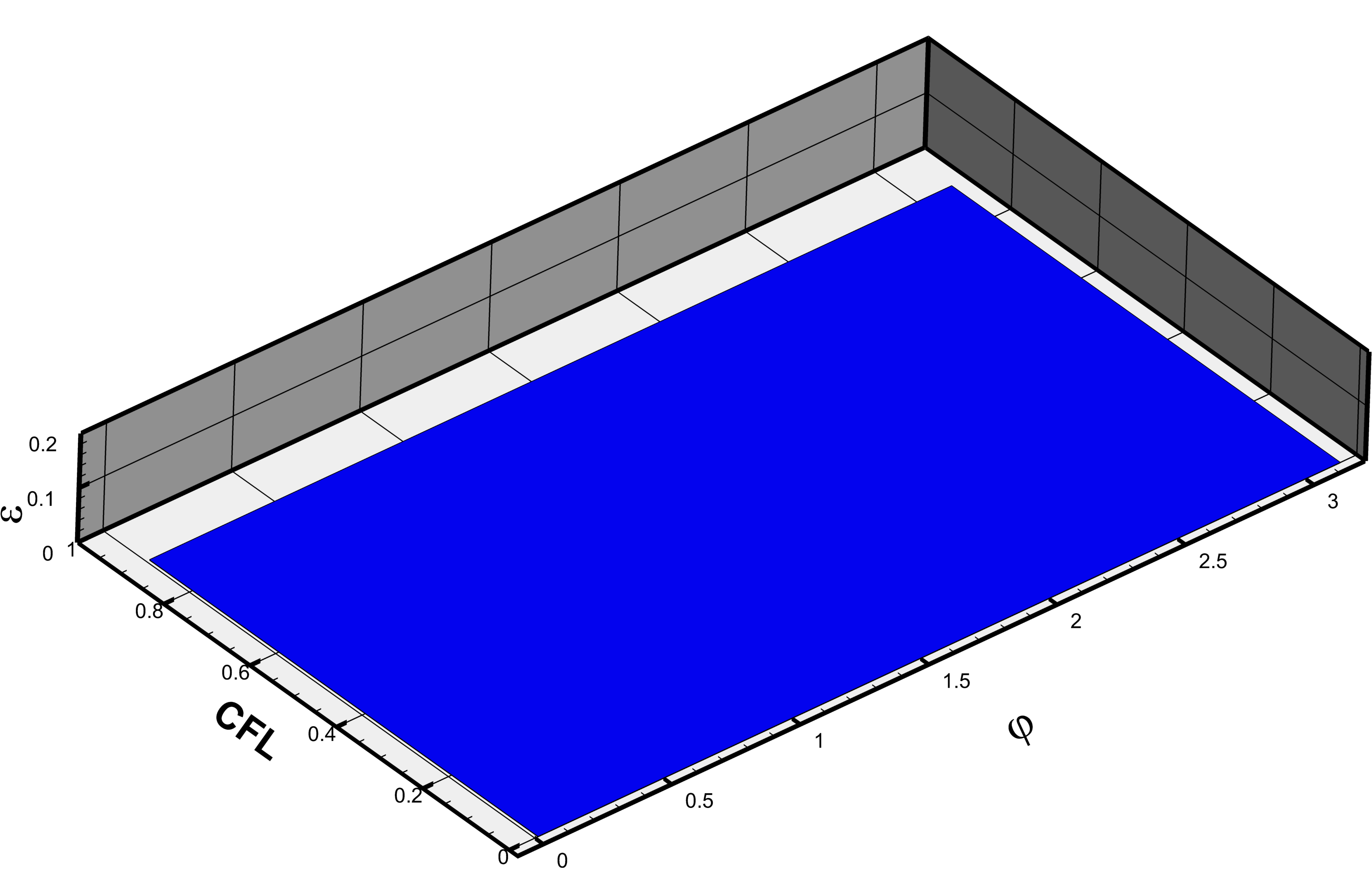}
 }
 \subfigure[\label{fig:ENO5}{ENO5}]{
 \includegraphics[width=0.48\textwidth]{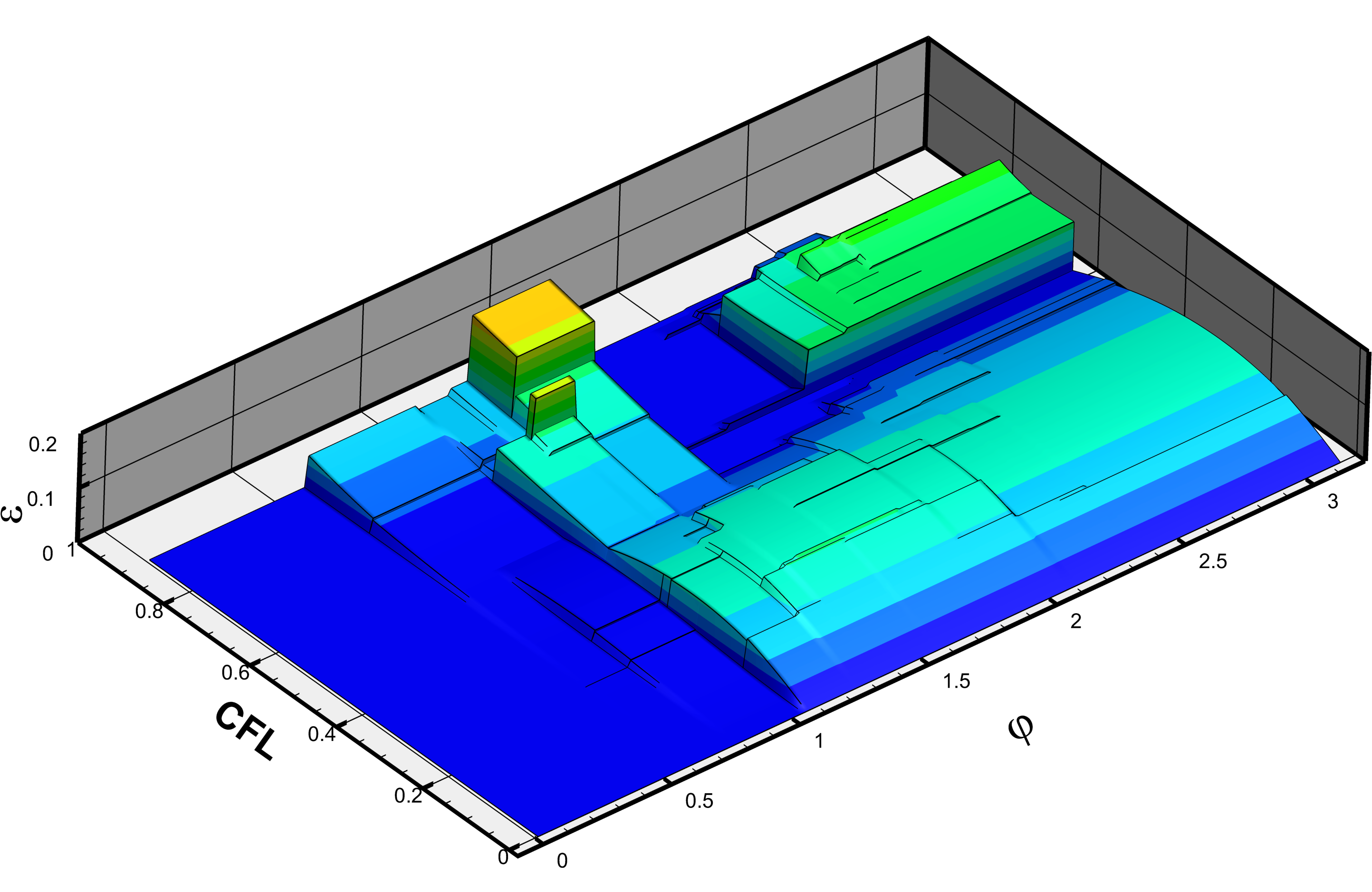}
 }
 \caption{ Numerical overshoots of ENO schemes.}
 \label{fig:ENO}
\end{figure}

It is encouraging and surprising to find that the third-order ENO scheme (ENO3) yields overshoot-free solutions, as shown in Fig.~\ref{fig:ENO3}. Whereas, the fifth-order ENO scheme (ENO5) produces an evident error, as indicated in Fig.~\ref{fig:ENO5}. This is expected since the five-point candidate stencils of ENO5 inevitably cross rather closely located discontinuities, and it is clearly demonstrated in Fig.~\ref{fig:ENO5} that the overshoot is directly related to the reduced wavenumber. As the reduced wavenumber goes above $1.05$, the overshoot error immediately increases regardless of the CFL number.
However, the CFL number also has an important effect on the overshoot error. For instance, as the CFL number is relatively large, the numerical overshoot is observable even in the low-wavenumber region ($\varphi<1.05$).
Another interesting finding is that the overshoot error is not monotonously increased with the CFL number, which is also  observed in the following results.

\begin{figure}[htbp]
 \centering
 \subfigure[\label{fig:WENO-JS5}{WENO-JS5}]{
 \includegraphics[width=0.48\textwidth]{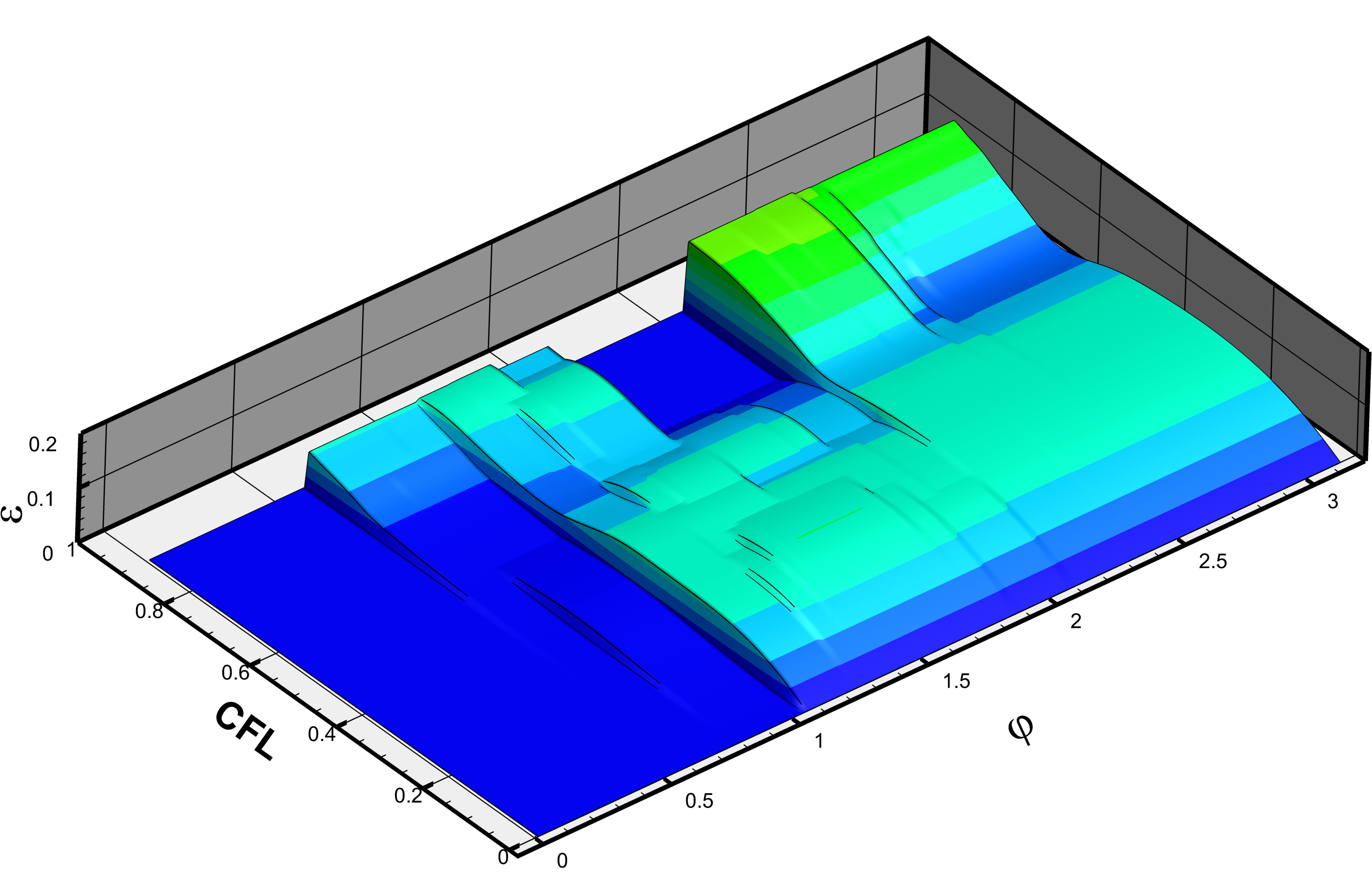}
 }
 \subfigure[\label{fig:WENO-JS7}{WENO-JS7}]{
 \includegraphics[width=0.48\textwidth]{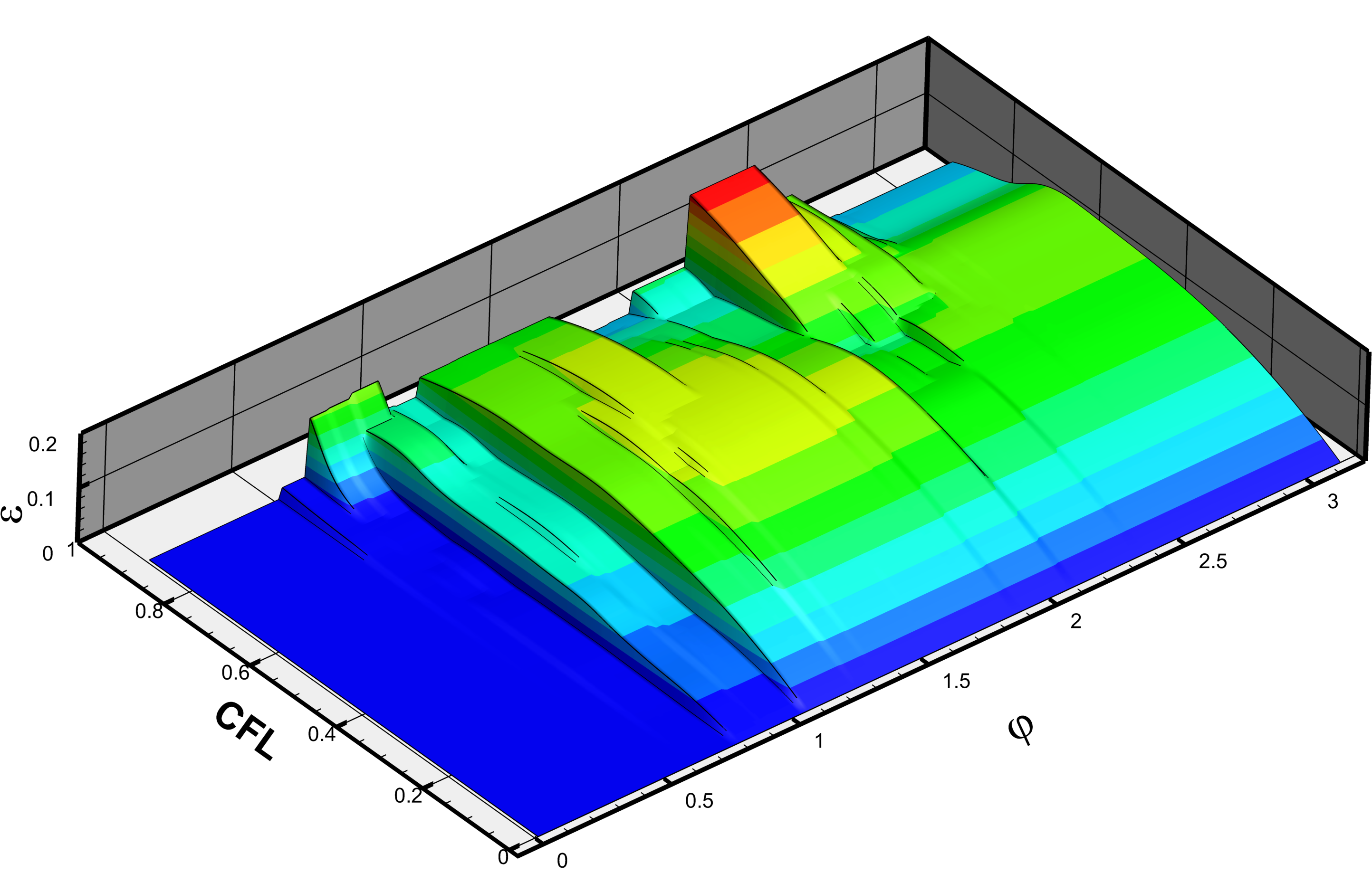}
 }
 \caption{Numerical overshoots of WENO-JS schemes.}
 \label{fig:WENO-JS}
\end{figure}

The result of the WENO-JS5 scheme in Fig.~\ref{fig:WENO-JS5} is similar to that of ENO5. As long as the discontinuities are loosely separated by a sufficiently large distance, namely, the reduced wavenumber is low, the smoothness of the candidate stencils can be guaranteed. As a result, the overshoot error is rather small, even when a large CFL number is used. Moreover, the overshoot error also does not monotonously vary with the CFL number for the WENO-JS5 scheme.
The WENO-JS7 scheme, in general, shows a similar behaviour, as found in Fig.\ref{fig:WENO-JS7}. However, the WENO-JS7 scheme already shows a noticeable overshoot as the reduced wavenumber is above $0.8$. This is believed to occur due to the fact that the candidate stencil in a relatively large width is prone to  cross    a discontinuity. It is also found that the amplitude of the overshoot produced by the WENO-JS7 scheme is significantly larger than that of the WENO-JS5 scheme.

\begin{figure}[htbp]
 \centering
 \subfigure[\label{fig:WENO-Z5}{WENO-Z5}]{
 \includegraphics[width=0.48\textwidth]{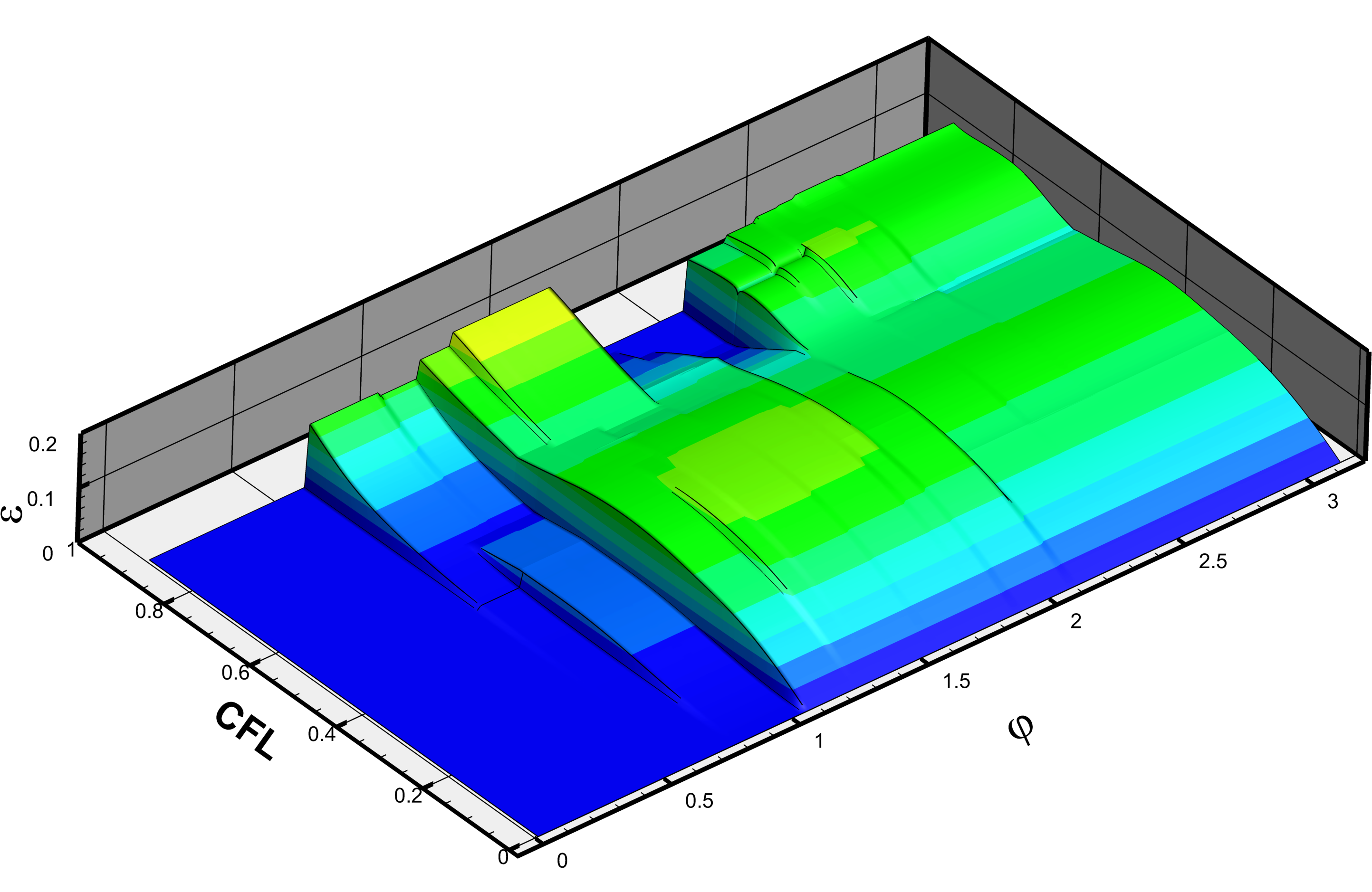}
 }
 \subfigure[\label{fig:WENO-Z7}{WENO-Z7}]{
 \includegraphics[width=0.48\textwidth]{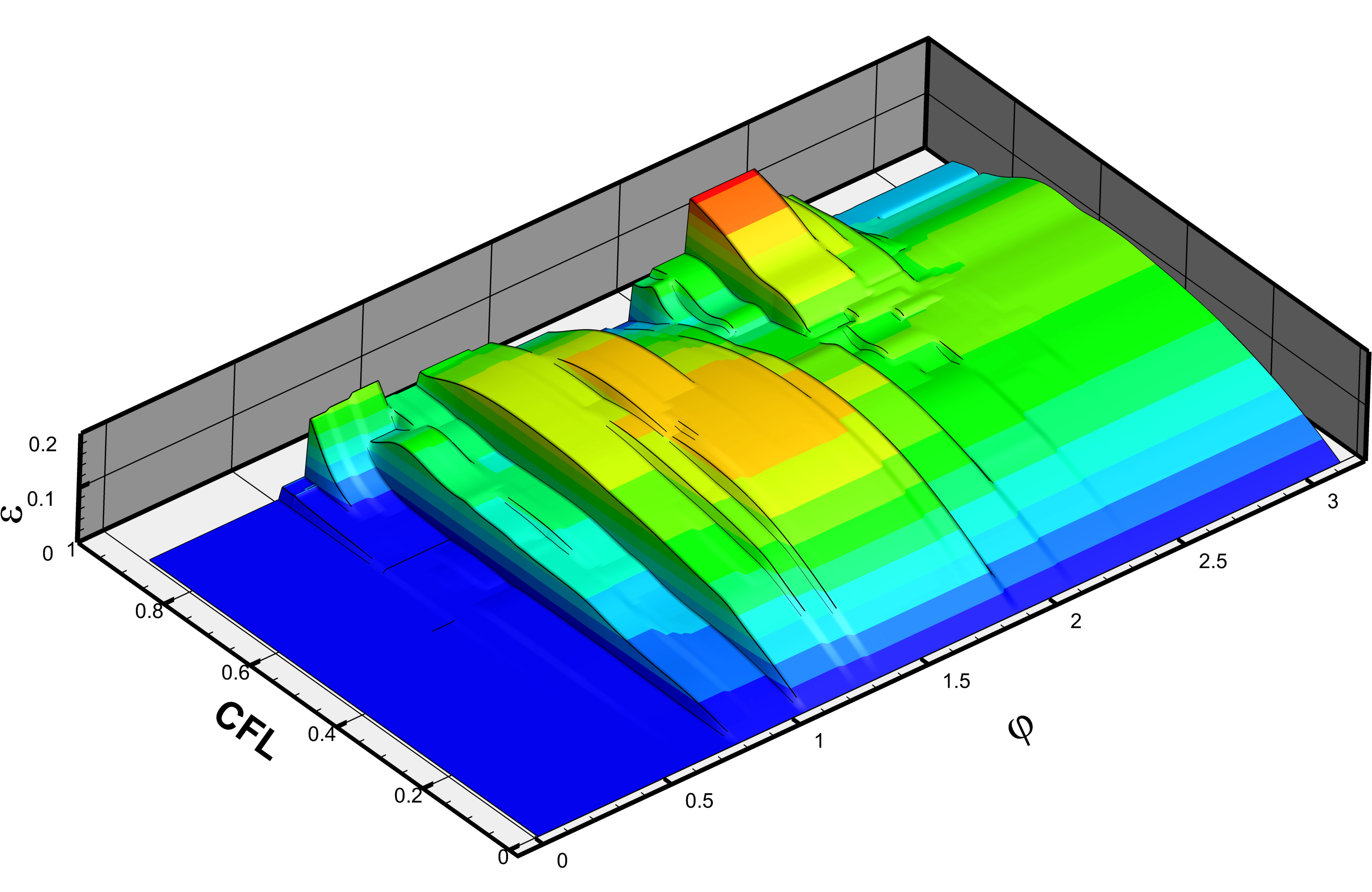}
 }
 \caption{Numerical overshoots of WENO-Z schemes.}
 \label{fig:WENO-Z}
\end{figure}

Fig.~\ref{fig:WENO-Z} shows the results of the WENO-Z schemes, which exhibit a similar pattern as for the WENO-JS schemes. However, the WENO-Z schemes, which are proved to be less dissipative compared with the WENO-JS schemes~\cite{Don2013}, produce more overshoots.
In particular, the overshoots in the intermediate wavenumber region are significant, compared against those of the WENO-JS schemes.

\begin{figure}[htbp]
 \centering
 \subfigure[\label{fig:MPWENO5}{MPWENO5}]{
 \includegraphics[width=0.48\textwidth]{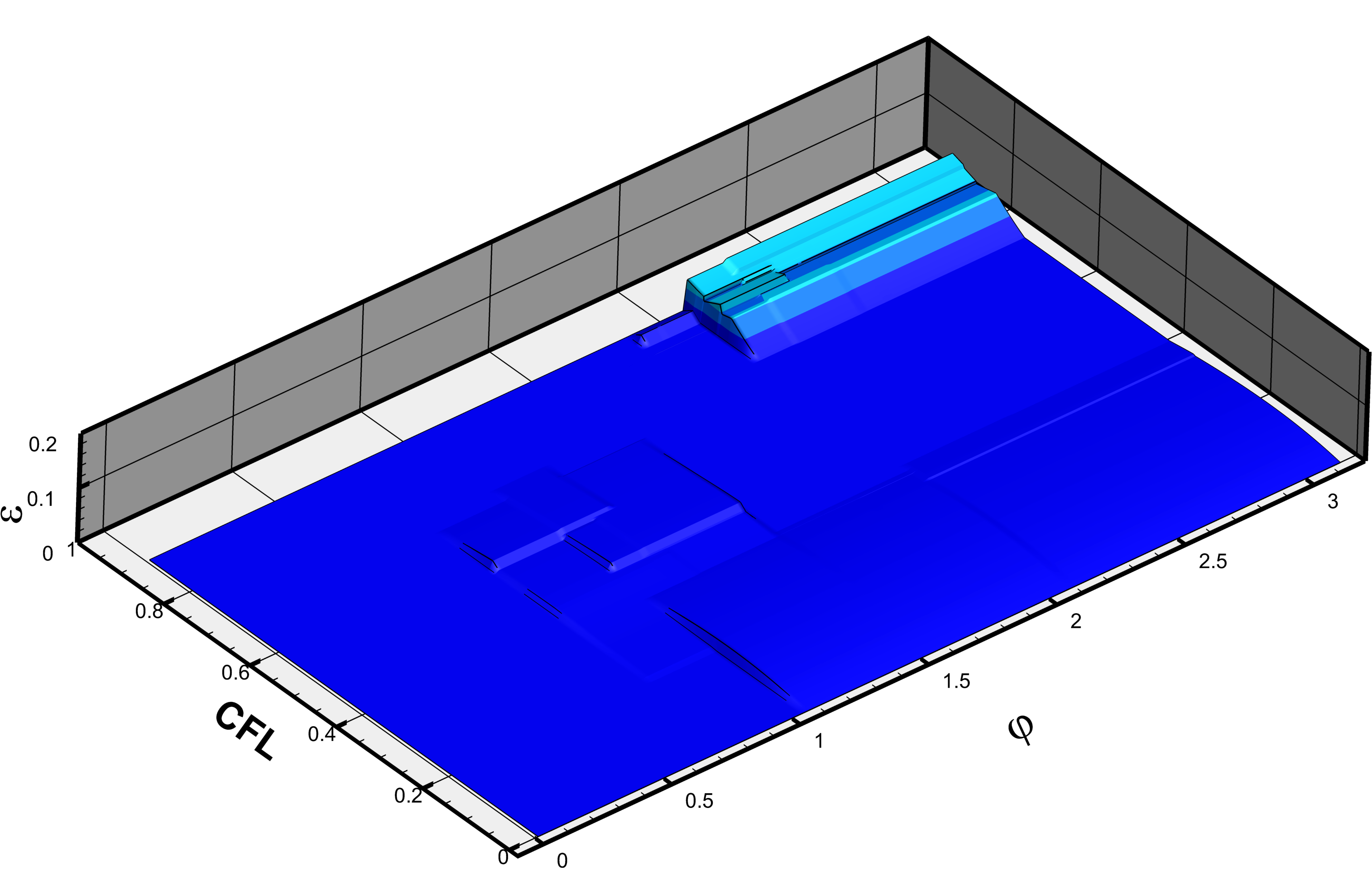}
 }
 \subfigure[\label{fig:MPWENO7}{MPWENO7}]{
 \includegraphics[width=0.48\textwidth]{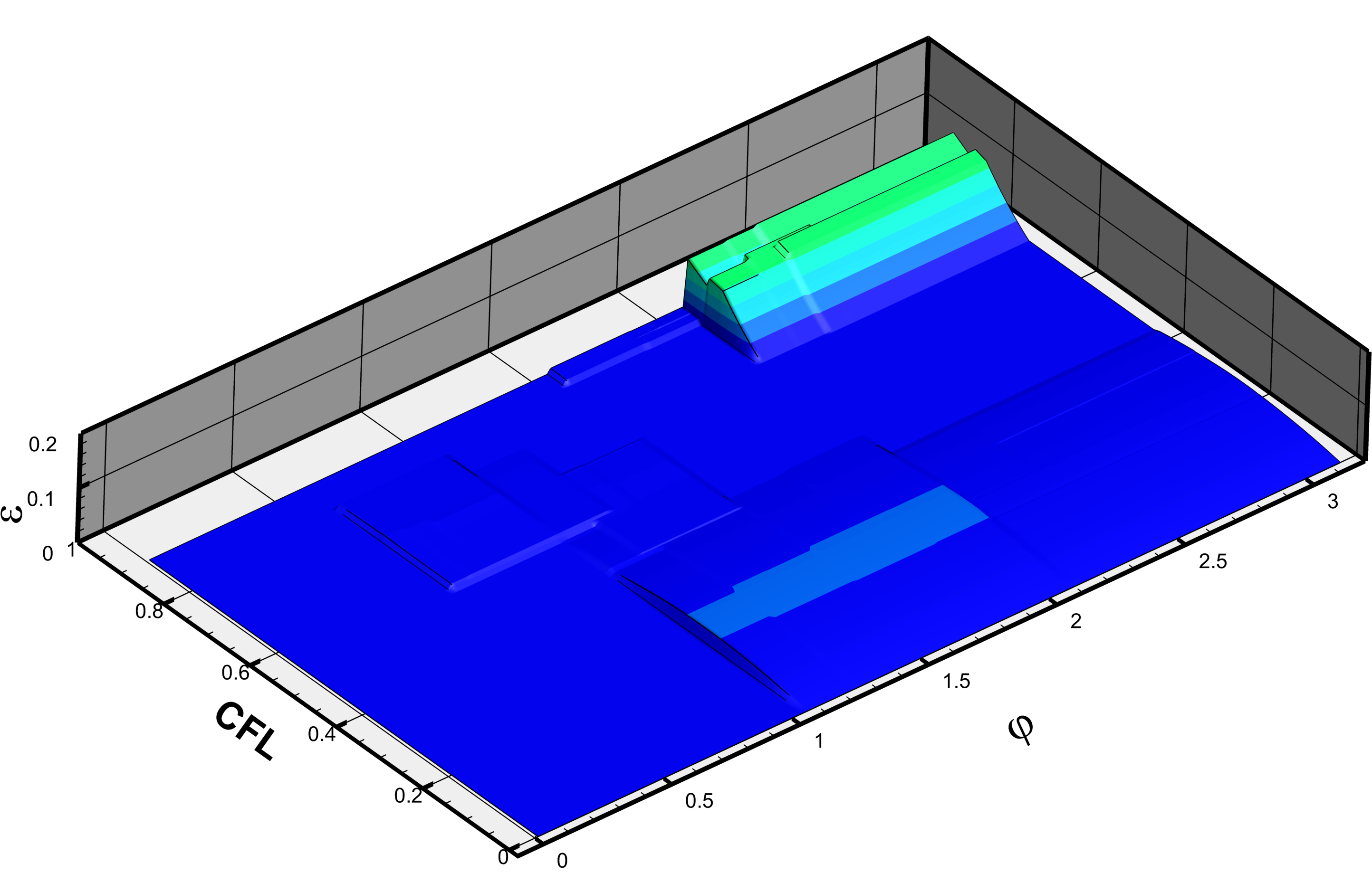}
 }
 \caption{ Numerical overshoots of MPWENO schemes.}
 \label{fig:MPWENO}
\end{figure}

The MPWENO schemes \cite{Balsara2000}, which incorporate the monotonicity-preserving bounds \cite{Suresh1997}, have significantly reduced the overshoot error, as shown in Fig.~\ref{fig:MPWENO}. The MPWENO7 scheme, having a formal order of accuracy of seven, only yields a relatively small overshoot even in the high-wavenumber region. Therefore, it is very effective to use the monotonicity-preserving bounds for improving the shock-capturing capability of WENO schemes.

\begin{figure}[htbp]
 \centering
 \subfigure[\label{fig:CFLInf}{$L_{\infty}$}]{
 \includegraphics[width=0.48\textwidth]{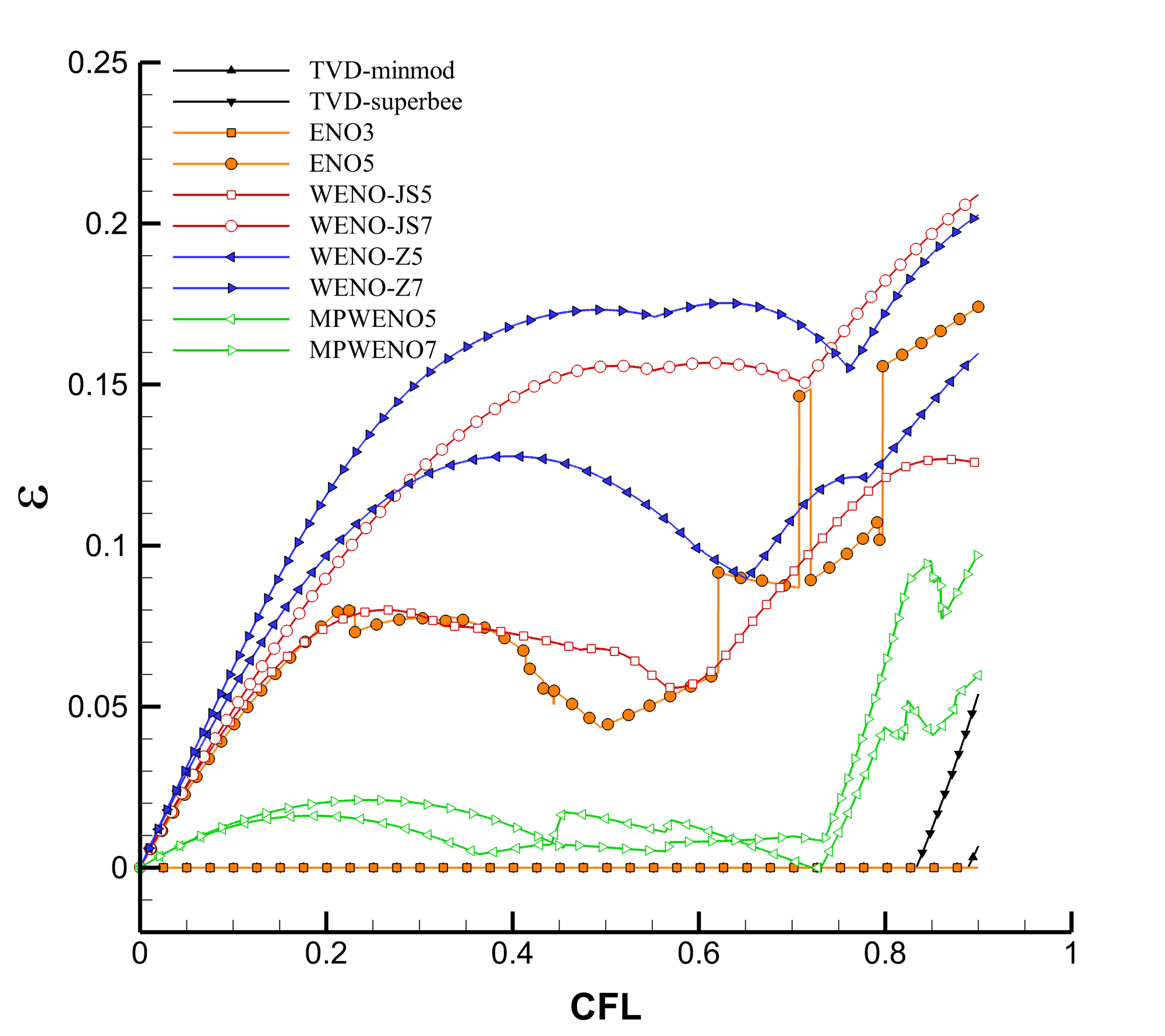}
 }
 \subfigure[\label{fig:CFL1}{$L_1$}]{
 \includegraphics[width=0.48\textwidth]{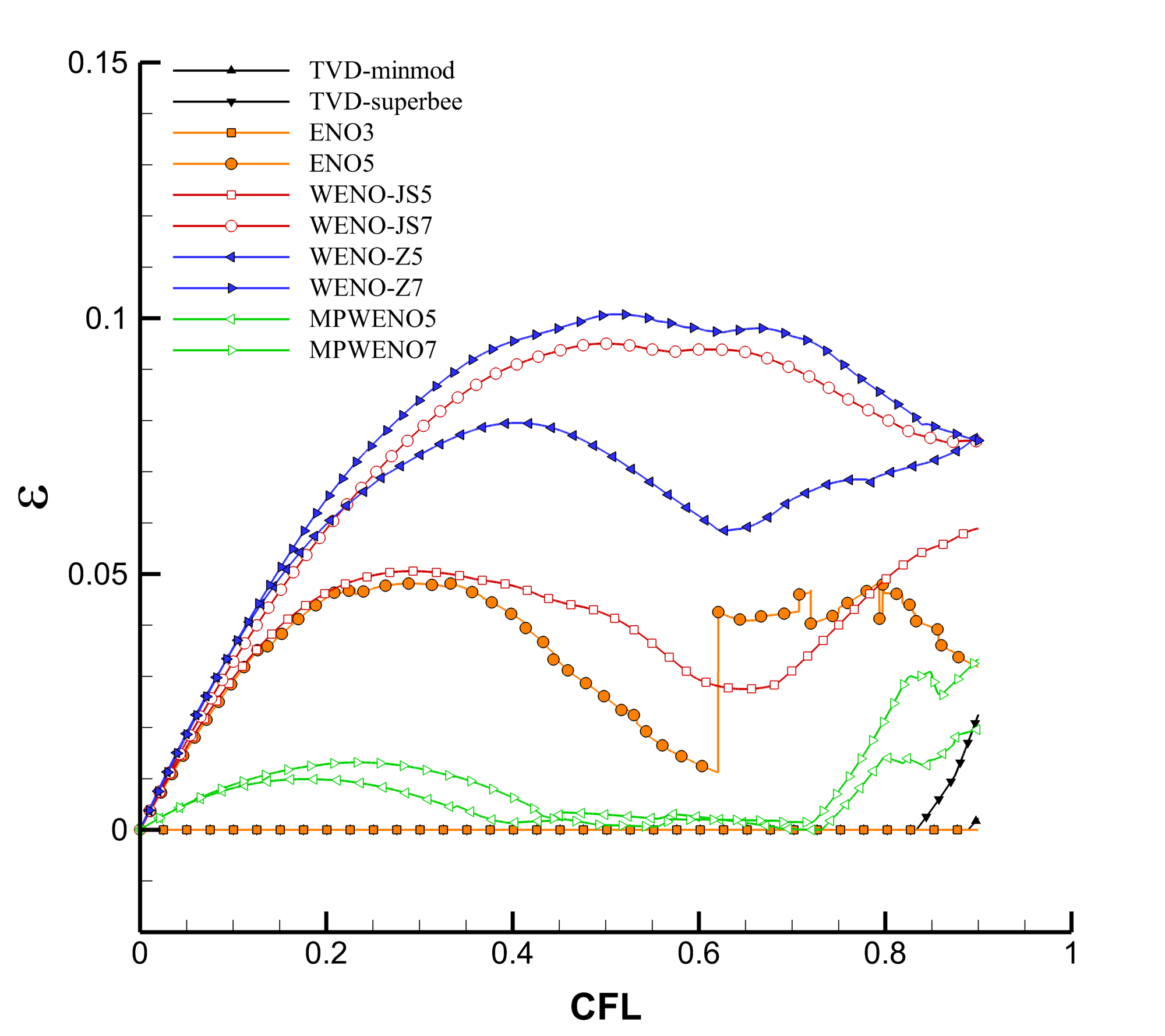}
 }
 \caption{ The overshoot  variations in the CFL number space.}
 \label{fig:CFL}
\end{figure}

\begin{figure}[htbp]
 \centering
 \subfigure[\label{fig:PhiInf}{$L_{\infty}$}]{
 \includegraphics[width=0.48\textwidth]{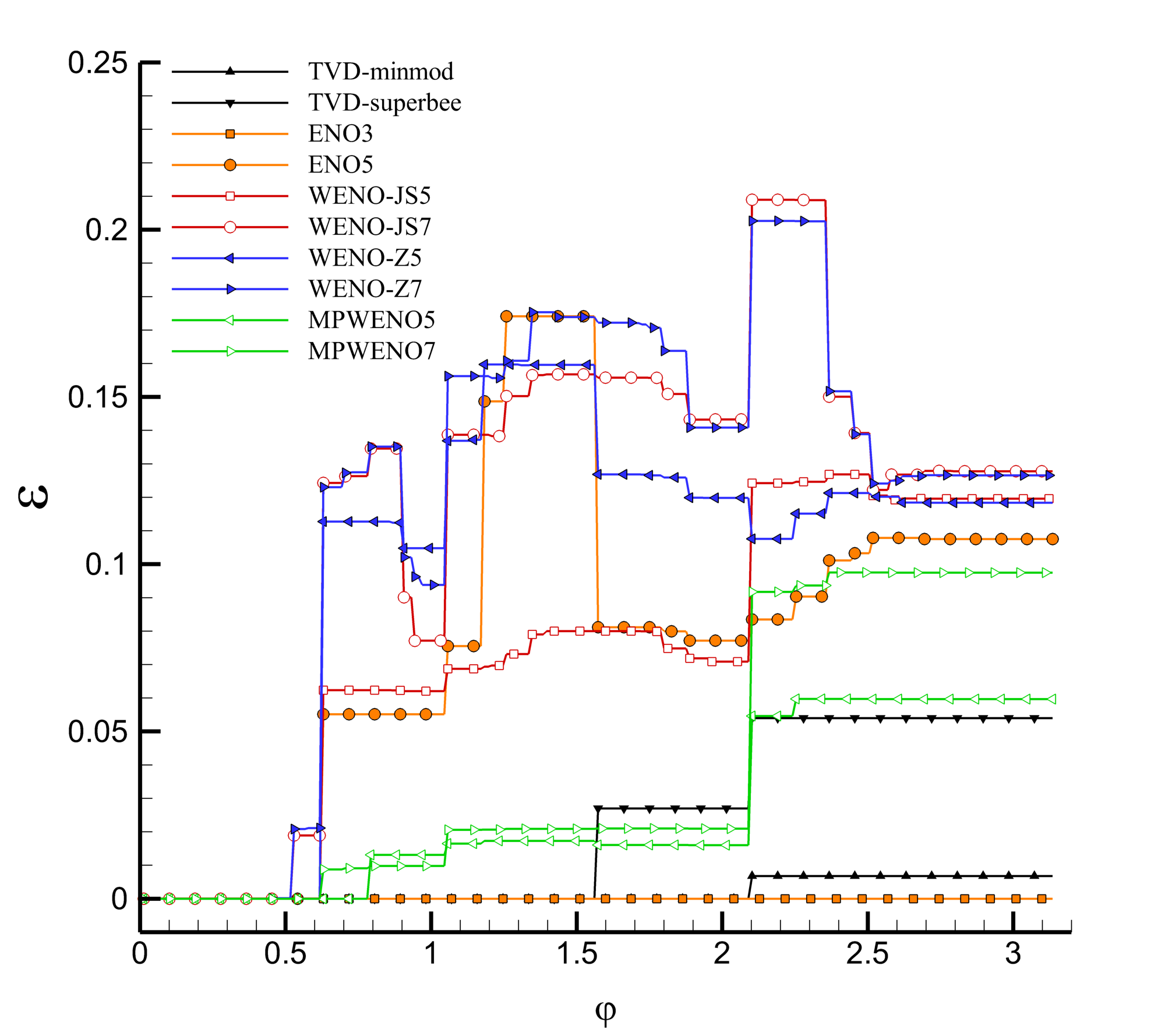}
 }
 \subfigure[\label{fig:Phi1}{$L_1$}]{
 \includegraphics[width=0.48\textwidth]{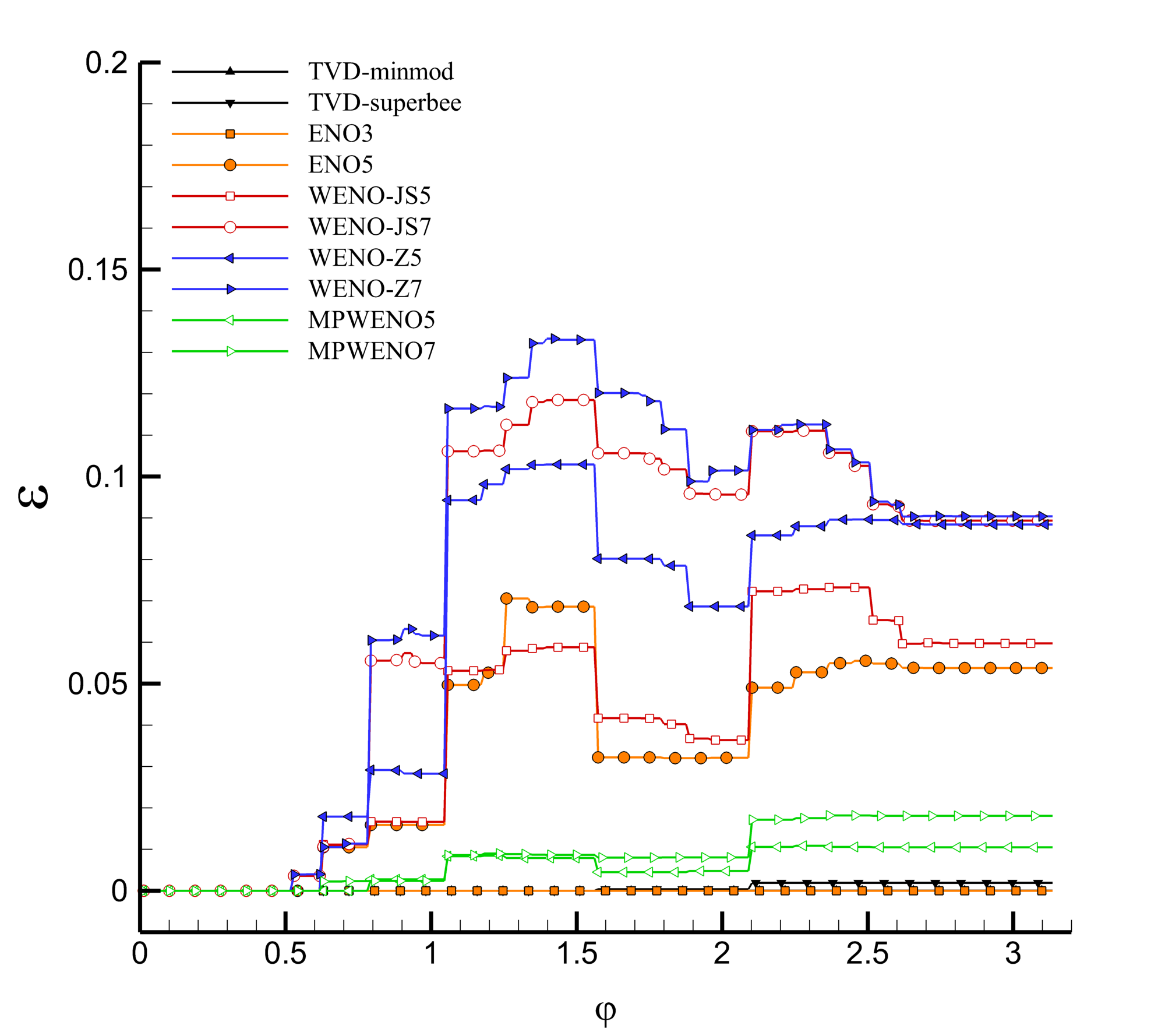}
 }
 \caption{ The overshoot  variations in the wavenumber space.}
 \label{fig:Phi}
\end{figure}

More details are shown in Fig.~\ref{fig:CFL} and Fig.~\ref{fig:Phi} in terms of the error norm. In these two figures, the previous results are projected into the CFL number space and the wavenumber space, respectively.

As shown in Fig.~\ref{fig:CFL}, the overshoot error norm behaves in a non-monotonic manner with respective to the CFL number. The most interesting behavior shown is that for several schemes relative small overshoots are obtained when the CFL number is close to 0.6.
In particular, the $L_{\infty}$ error norm of the WENO-Z5 scheme is reduced by $30\%$ with the CFL number increased from $0.4$ to $0.65$.

In Fig.~\ref{fig:Phi}, the error norm exhibits as a sharp staircase function. The TVD schemes with the SB and the MM limiters perform well in suppressing the overshoot within the whole region of the reduced wavenumber, and we have found again that the most robust high-order shock-capturing schemes are the MPWENO schemes by comparing the error norms among the schemes considered.
All schemes are capable of producing overshoot-free results in the region of low reduced wavenumbers. WENO-type schemes start to produce significant overshoots as the reduced wavenumber is greater than 1.05.

Moreover, while for the TVD schemes and the MPWENO schemes the $L_\infty$ error norm is a monotonic function of the reduced wavenumer, it is not for the other schemes. Take WENO-JS5 for instance, the error norm rises and falls as the reduced wavenumer is increased from 1 to 3. We leave the further investigation of the underlying mechanism associated to WENO-JS5 in this case, as well as the specific characteristic of each of the other schemes in the future work.

\begin{figure}[htbp]
 \centering
 \subfigure[\label{fig:TVDandENO}{TVD and ENO}]{
 \includegraphics[width=0.48\textwidth]{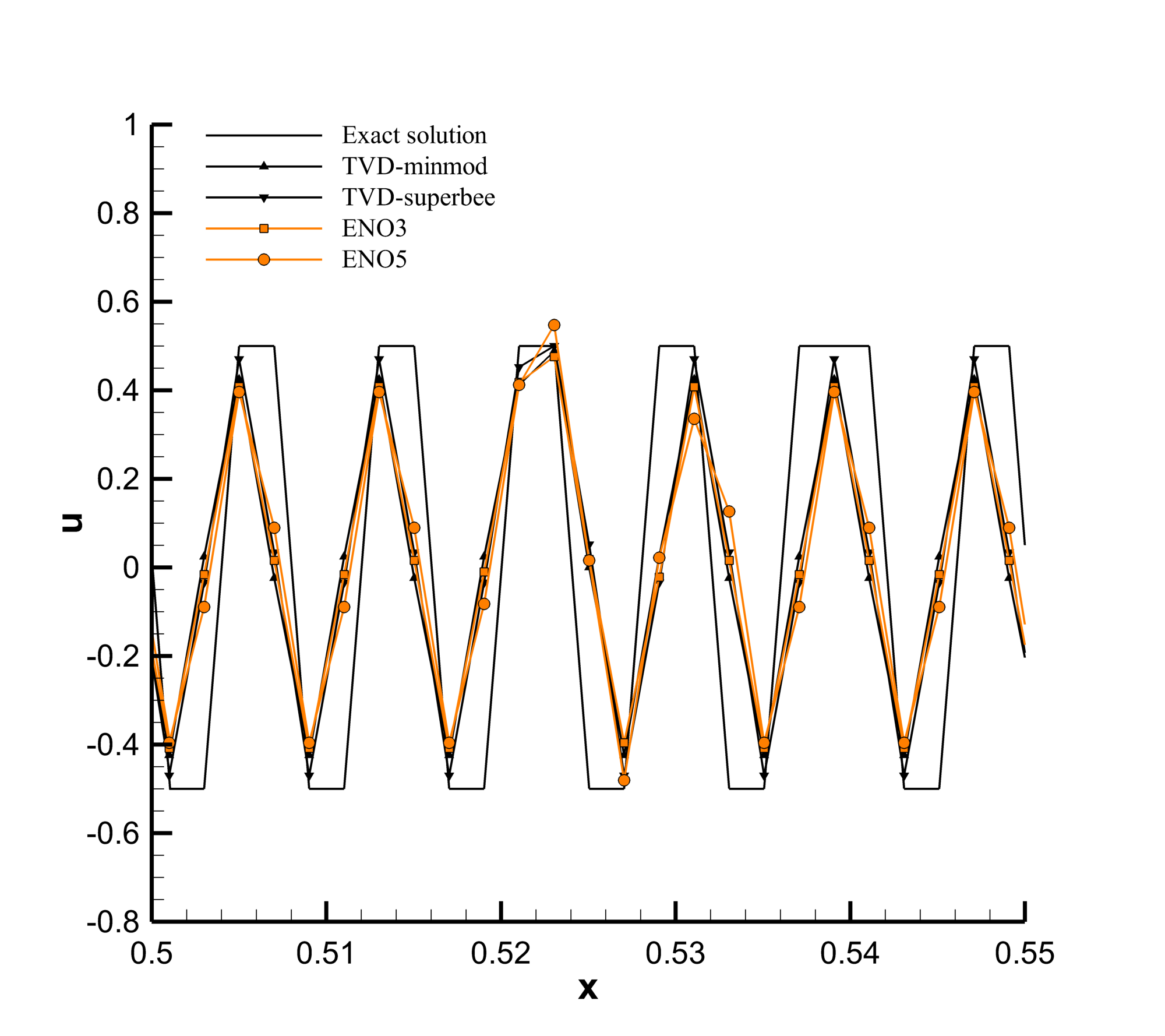}
 }
 \subfigure[\label{fig:WENOS}{WENO}]{
 \includegraphics[width=0.48\textwidth]{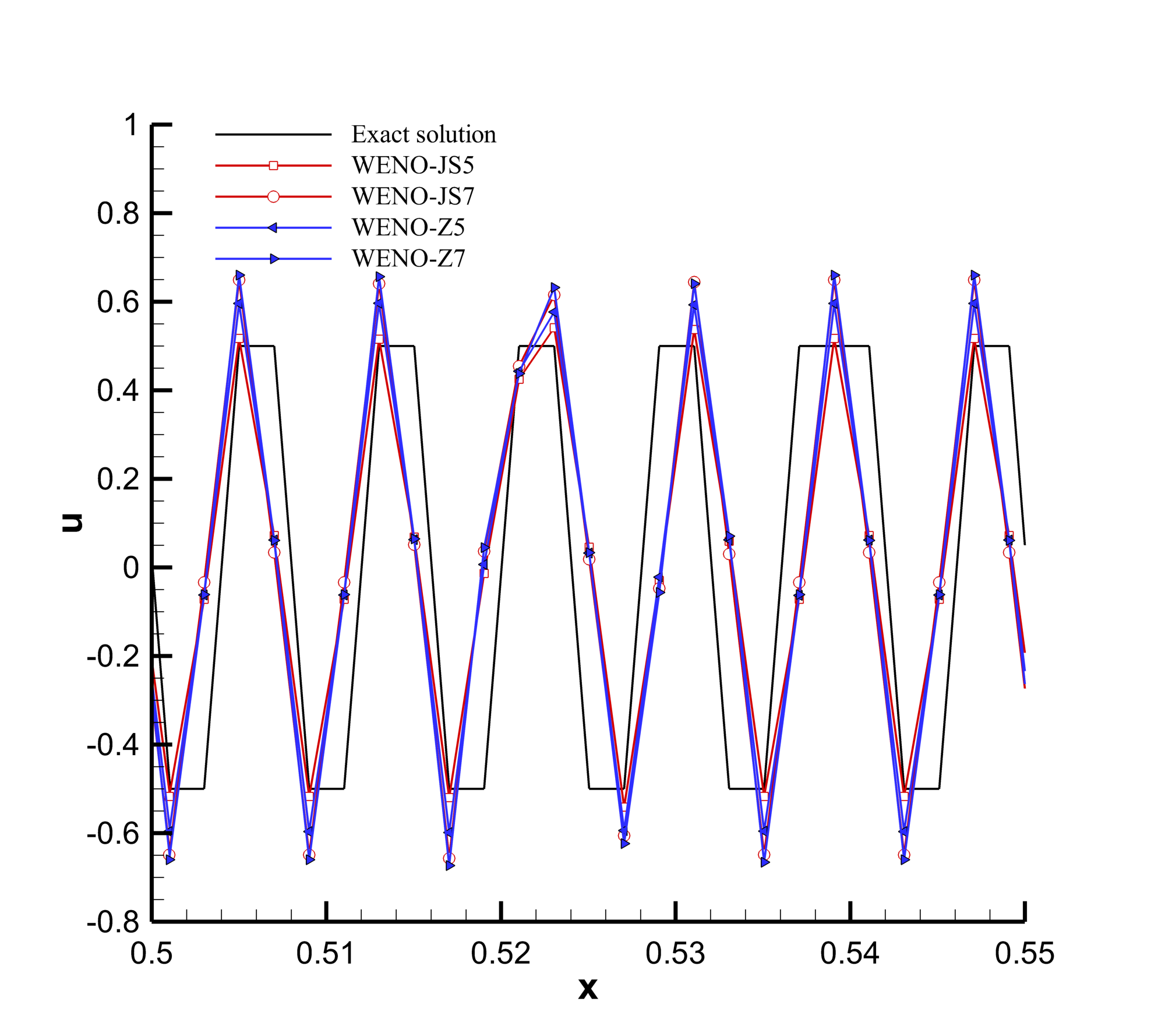}
 }
  \subfigure[\label{fig:MPWENOS}{MPWENO}]{
 \includegraphics[width=0.48\textwidth]{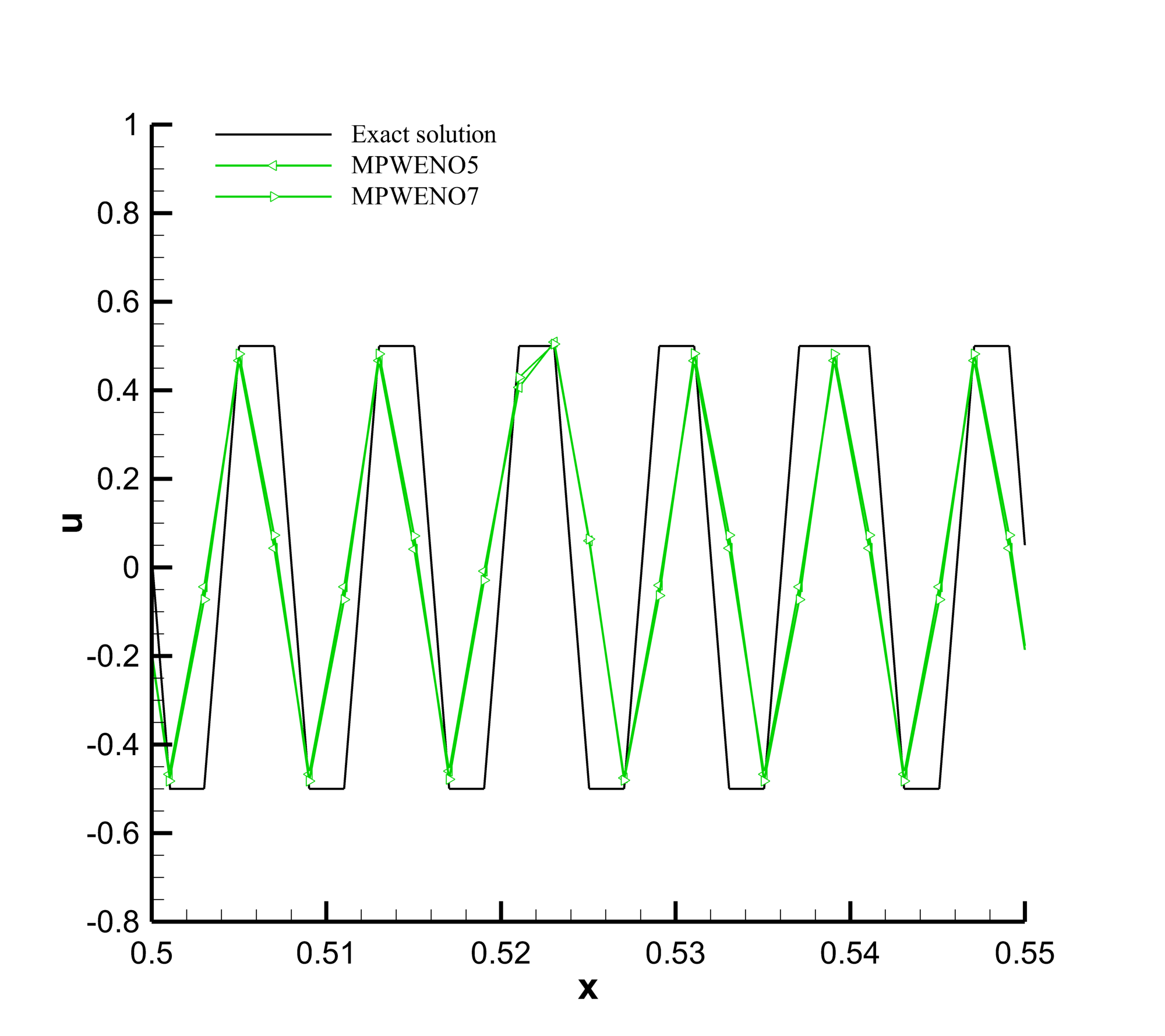}
 }
 \caption{ The overshoots distribution zoomed in space for CFL=0.6 and $\varphi$=1.5.}
 \label{fig:OVERSHOOT}
\end{figure}

Finally, in Fig.~\ref{fig:OVERSHOOT}, we provide the distribution of the overshoot zoomed in space $x\in [0.5,0.55]$ for CFL$=0.6$ and $\varphi$=1.5. Note that due to the limited resolution, the exact solution is not really a square wave train. We can see that the solutions of the TVD schemes, the ENO3 scheme, and the MPWENO schemes are overshoot-free. The WENO-JS5 scheme is also relatively robust, but the other WENO schemes, especially the higher-order ones, produce strong overshoots.

\section{Conclusions}\label{sec:Conclusions}
We have designed a simple numerical metric for the evaluation of overshoot errors yielded by a limited number of representative shock-capturing schemes.
Despite that the method introduced only involves discontinuities, and smooth waves are not considered, one may still use it to quantitatively evaluate the shock-capturing capability of the numerical schemes.
According to the obtained results, several conclusions can be drawn straightforwardly: (I) the amplitude of overshoots significantly depend on the reduced wavenumber, which indicates the distance of closely located discontinuities; (II) the amplitude of overshoots does not monotonously vary with the CFL number; and (III) the maximum overshoots of different schemes may be produced in very different scenarios, concerning the combination of CFL number and the reduced wavenumber.

Since the analysis is based on a quantitative evaluation, the present method goes beyond the common case-by-case fashion, which
mainly relies on the visualized investigation. In particular, in a complex and more realistic simulation, it is difficult to observe
the exact cause of numerical oscillations. Therefore, we expect the present method to be used as an effective tool for the evaluation of new shock-capturing schemes, some of which may be used to address critical situations.

It should be noted that the magnitude of the overshoot error may vary a bit while using a different {time-stepping technique}, but it does not affect the aforementioned conclusions from our investigation.

\section*{Acknowledgment}
This work was partially supported by the National Key Project (Grant No. GJXM92579).

\bibliography{ref}

\begin{thebibliography}{10}
\expandafter\ifx\csname url\endcsname\relax
  \def\url#1{\texttt{#1}}\fi
\expandafter\ifx\csname urlprefix\endcsname\relax\def\urlprefix{URL }\fi
\expandafter\ifx\csname href\endcsname\relax
  \def\href#1#2{#2} \def\path#1{#1}\fi

\bibitem{Harten1984}
A.~Harten, On a class of high resolution total-variation-stable
  finite-difference schemes, SIAM J. Numer. Anal. 21~(1) (1984) 1--23.
\newblock \href {http://dx.doi.org/10.1137/0721001}
  {\path{doi:10.1137/0721001}}.

\bibitem{Harten1987}
A.~Harten, B.~Engquist, S.~Osher, S.~R. Chakravarthy, Uniformly high order
  accurate essentially non-oscillatory schemes, {III}, Journal of Computational
  Physics 71~(2) (1987) 231--303.
\newblock \href {http://dx.doi.org/10.1016/0021-9991(87)90031-3}
  {\path{doi:10.1016/0021-9991(87)90031-3}}.

\bibitem{Liu1994}
X.~D. Liu, S.~Osher, T.~Chan, Weighted essentially non-oscillatory schemes,
  Journal of Computational Physics 115~(1) (1994) 200--212.
\newblock \href {http://dx.doi.org/10.1006/jcph.1994.1187}
  {\path{doi:10.1006/jcph.1994.1187}}.

\bibitem{Zhang2020a}
H.~Zhang, F.~Zhang, J.~Liu, J.~McDonough, C.~Xu, A simple extended compact
  nonlinear scheme with adaptive dissipation control, Communications in
  Nonlinear Science and Numerical Simulation 84 (2020) 105191.
\newblock \href {http://dx.doi.org/10.1016/j.cnsns.2020.105191}
  {\path{doi:10.1016/j.cnsns.2020.105191}}.

\bibitem{Xu2021}
C.~Xu, F.~Zhang, H.~Dong, H.~Jiang, Arbitrary high-order extended essentially
  non-oscillatory schemes for hyperbolic conservation laws, International
  Journal for Numerical Methods in Fluids n/a~(n/a).
\newblock \href {http://dx.doi.org/10.1002/fld.4968}
  {\path{doi:10.1002/fld.4968}}.

\bibitem{Jiang1996}
G.-S. Jiang, C.-W. Shu, Efficient implementation of weighted {ENO} schemes,
  Journal of Computational Physics 126~(1) (1996) 202--228.
\newblock \href {http://dx.doi.org/10.1006/jcph.1996.0130}
  {\path{doi:10.1006/jcph.1996.0130}}.

\bibitem{Borges2008}
R.~Borges, M.~Carmona, B.~Costa, W.~S. Don, An improved weighted essentially
  non-oscillatory scheme for hyperbolic conservation laws, Journal of
  Computational Physics 227~(6) (2008) 3191--3211.
\newblock \href {http://dx.doi.org/10.1016/j.jcp.2007.11.038}
  {\path{doi:10.1016/j.jcp.2007.11.038}}.

\bibitem{Zhao2019}
G.~Zhao, M.~Sun, A.~Memmolo, S.~Pirozzoli, A general framework for the
  evaluation of shock-capturing schemes, Journal of Computational Physics 376
  (2019) 924--936.
\newblock \href {http://dx.doi.org/10.1016/j.jcp.2018.10.013}
  {\path{doi:10.1016/j.jcp.2018.10.013}}.

\bibitem{Balsara2000}
D.~S. Balsara, C.-W. Shu, Monotonicity preserving weighted essentially
  non-oscillatory schemes with increasingly high order of accuracy, Journal of
  Computational Physics 160~(2) (2000) 405 -- 452.
\newblock \href {http://dx.doi.org/10.1006/jcph.2000.6443}
  {\path{doi:10.1006/jcph.2000.6443}}.

\bibitem{Hu2013}
X.~Y. Hu, N.~A. Adams, C.-W. Shu, Positivity-preserving method for high-order
  conservative schemes solving compressible euler equations, Journal of
  Computational Physics 242 (2013) 169 -- 180.
\newblock \href {http://dx.doi.org/https://doi.org/10.1016/j.jcp.2013.01.024}
  {\path{doi:https://doi.org/10.1016/j.jcp.2013.01.024}}.

\bibitem{Sweby1984}
P.~Sweby, High resolution schemes using flux limiters for hyperbolic
  conservation laws, SIAM J. Numer. Anal. 21~(5) (1984) 995--1011.
\newblock \href {http://dx.doi.org/10.1137/0721062}
  {\path{doi:10.1137/0721062}}.

\bibitem{Deng2019}
X.~Deng, Y.~Shimizu, F.~Xiao, A fifth-order shock capturing scheme with
  two-stage boundary variation diminishing algorithm, Journal of Computational
  Physics 386 (2019) 323--349.
\newblock \href {http://dx.doi.org/10.1016/j.jcp.2019.02.024}
  {\path{doi:10.1016/j.jcp.2019.02.024}}.

\bibitem{Gottlieb2001}
S.~Gottlieb, C.~Shu, E.~Tadmor, Strong stability-preserving high-order time
  discretization methods, SIAM Review 43~(1) (2001) 89--112.
\newblock \href {http://dx.doi.org/10.1137/S003614450036757X}
  {\path{doi:10.1137/S003614450036757X}}.

\bibitem{Osher1984}
S.~Osher, S.~Chakravarthy, High resolution schemes and the entropy condition,
  SIAM Journal on Numerical Analysis 21~(5) (1984) 955--984.
\newblock \href {http://arxiv.org/abs/https://doi.org/10.1137/0721060}
  {\path{arXiv:https://doi.org/10.1137/0721060}}, \href
  {http://dx.doi.org/10.1137/0721060} {\path{doi:10.1137/0721060}}.

\bibitem{Shu1989}
C.-W. Shu, S.~Osher, Efficient implementation of essentially non-oscillatory
  shock-capturing schemes,ii, J. Comput. Phys. 83~(1) (1989) 32--78.
\newblock \href {http://dx.doi.org/10.1016/0021-9991(89)90222-2}
  {\path{doi:10.1016/0021-9991(89)90222-2}}.

\bibitem{Don2013}
W.-S. Don, R.~Borges, Accuracy of the weighted essentially non-oscillatory
  conservative finite difference schemes, Journal of Computational Physics 250
  (2013) 347 -- 372.
\newblock \href {http://dx.doi.org/10.1016/j.jcp.2013.05.018}
  {\path{doi:10.1016/j.jcp.2013.05.018}}.

\bibitem{Suresh1997}
A.~Suresh, H.~Huynh, Accurate monotonicity-preserving schemes with
  runge–kutta time stepping, Journal of Computational Physics 136~(1) (1997)
  83 -- 99.
\newblock \href {http://dx.doi.org/10.1006/jcph.1997.5745}
  {\path{doi:10.1006/jcph.1997.5745}}.

\end{thebibliography}
\end{document}